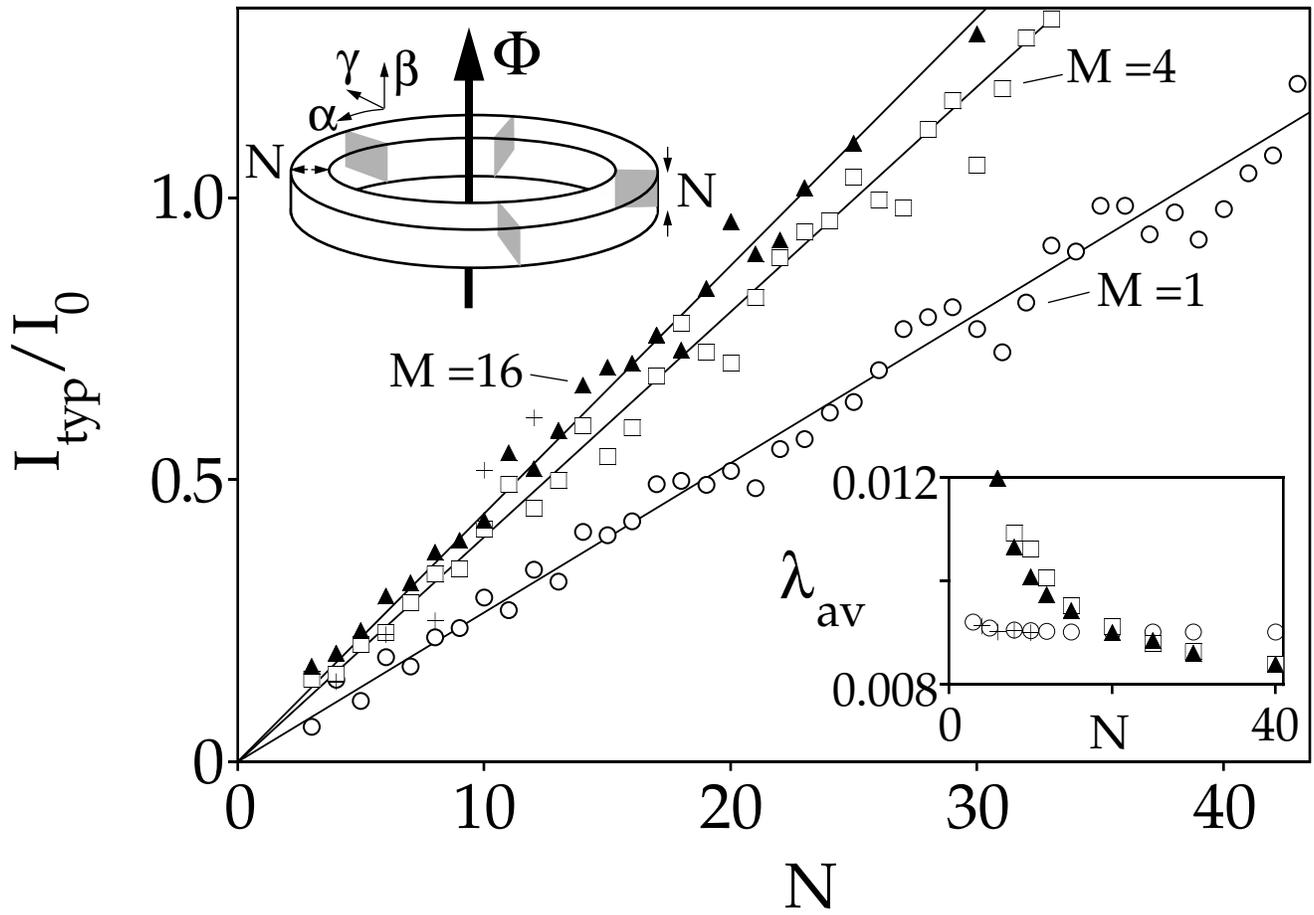

Figure 1

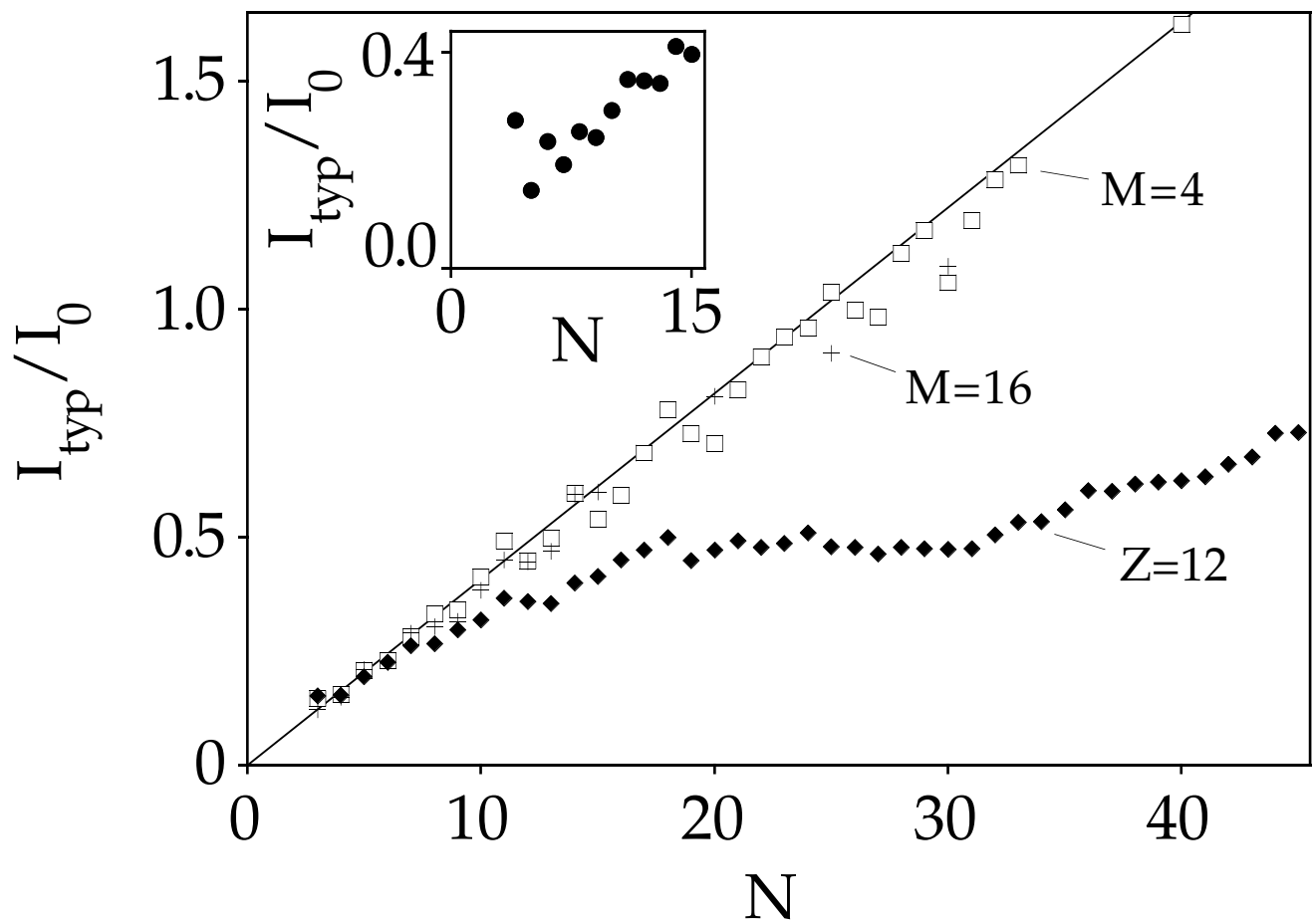

Figure 2

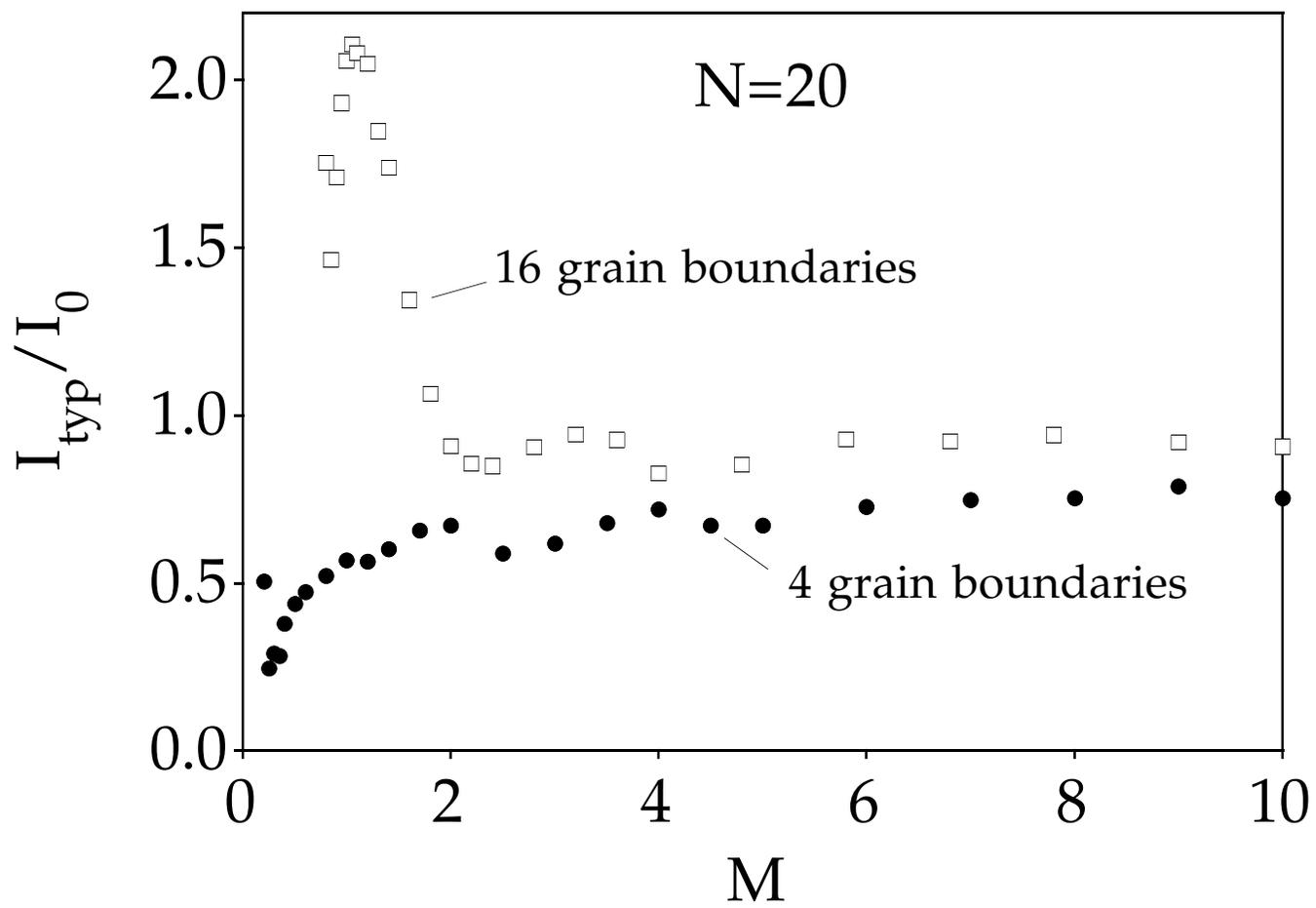

Figure 3

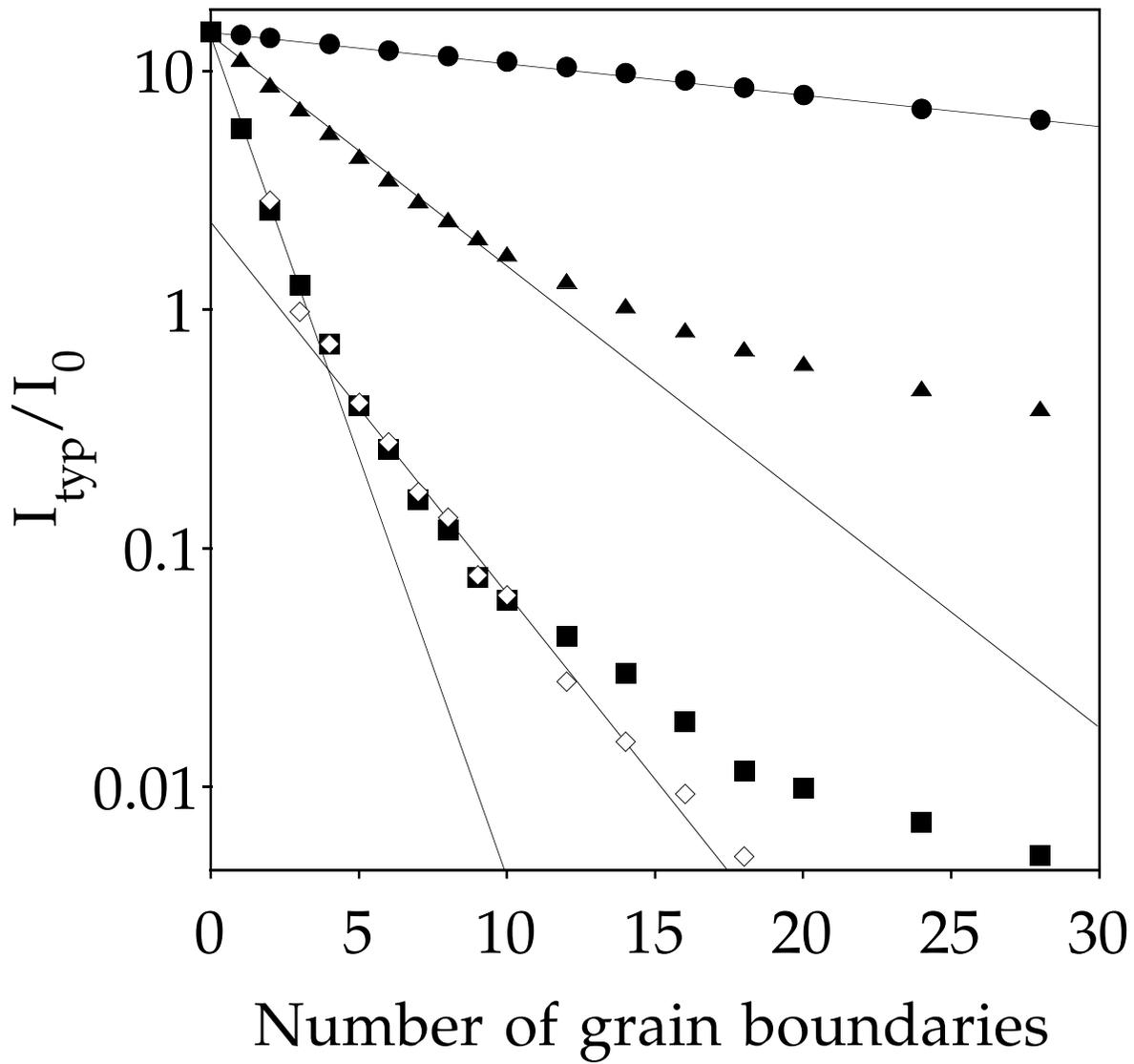

Figure 4

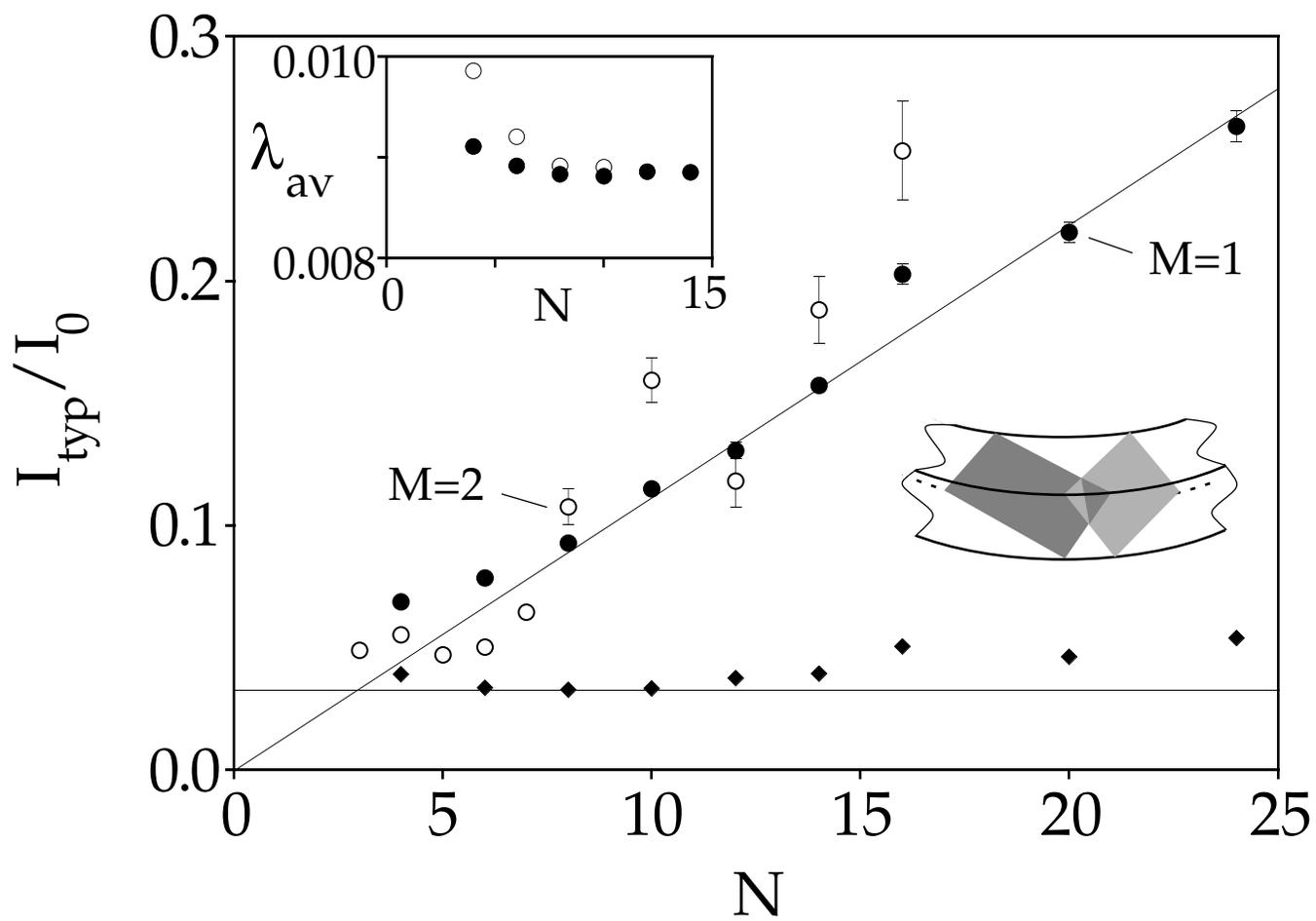

Figure 5

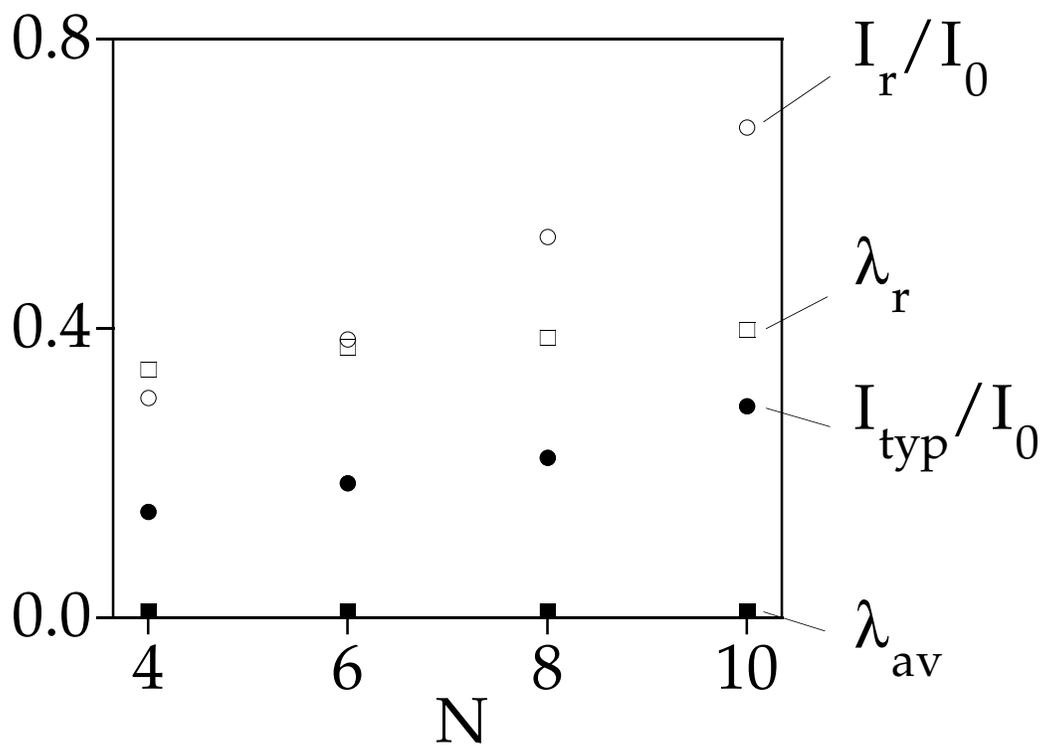

Figure 6

# Why Are Large Persistent Currents Observed in Small Gold Rings?

George Kirczenow

*Department of Physics, Simon Fraser University,*
*Burnaby, British Columbia,*
*Canada, V5A 1S6*

It is demonstrated using three-dimensional computer simulations that some simple non-interacting electron models that include electron scattering by grain boundaries, exhibit coexistence of large persistent currents and small conductances, similar to that observed experimentally in isolated micron-scale gold rings. Models with simple grain boundaries, and models with small numbers of regularly stepped or atomically rough dilute grain boundaries have been studied and found to yield similar results, which differ markedly, however, from the predictions of models that assume only random impurity scattering. This difference is due to the fact that equilibrium persistent currents and non-equilibrium transport coefficients are physically different things and depend in different ways on the *topology* of the defect structure in a conducting ring. Experiments on metal and semiconductor rings that should be able to determine whether this is the explanation of the effects observed by Chandrasekhar *et al*. (Phys. Rev. Lett. **67**, 3578 (1991)) are proposed.

PACS. 73.20.Dx − Electron states in low-dimensional structures.
PACS. 73.40.-c − Electronic transport in interface structures.
PACS. 73.50.Jt − Magnetotransport effects.

# 1. Introduction

When a conductor in its ground state is immersed in a magnetic field, it acquires a magnetic moment and an associated circulating electric current. This current, being an *equilibrium* property, does not dissipate. It is thus referred to as a "persistent current." The basic theory of persistent currents in normal metal rings was formulated by London[1] in the 1930's in the context of aromatic molecules. It was developed further by several authors in the 1960's.[2-7] More recently, Büttiker, Landauer and Imry[8] showed that these currents should occur even in the presence of disorder. Cheung, Riedel and Gefen[9] then estimated the magnitude of the persistent currents that should flow in small, disordered normal metal rings with non-interacting electrons. They predicted that, at low temperatures, the typical persistent current should be $I_{typ} \sim I_0 \, l/L$ where $I_0 = e v_F/L$, $e$ is the electron charge, $L$ is the ring's circumference, $v_F$ is the electron Fermi velocity, and $l$ is the elastic mean free path. Since then, persistent currents have been observed in an array of $10^7$ micron-scale copper rings by Lévy *et al*,[10] and in individual gold and semiconductor rings by Chandrasekhar *et al* [11] and Mailly *et al*.[12] The measurements on the single gold rings[11] presented a puzzle that has remained unresolved: The measured persistent currents were of order $I_0$ in rings for which $l/L \sim 0.01$, i.e., the observed currents were *two orders of magnitude larger* than had been predicted[9] for rings with such small transport mean free paths and correspondingly small conductances.

During the last three years, there have been several attempts to explain this observation as a novel effect of electron-electron interactions: Very large enhancements of the persistent current due to many-body effects were predicted in three-dimensional rings[13,14] and also in one and two-dimensional models.[15] However, subsequent more detailed three-dimensional calculations[16,17,18] yielded no significant many-body enhancement of the persistent current. Also exact results of Luttinger liquid theory for defect-free rings[19] as well as exact numerical solutions of models of small systems with and without defects[20-24] have indicated that electron-electron interactions do not yield any major enhancement of the persistent current in one-dimensional rings. A recent theoretical study of two interacting electrons in a narrow two-dimensional ring also showed no enhancement of the persistent current by electron-electron interactions.[25] Furthermore, the results of persistent current measurements on two-dimensional, few-channel, quasi-ballistic semiconductor rings[12] have been in reasonably good agreement with the predictions of the non-interacting electron theory,[9] so that it seems that there is no large many-body enhancement of the persistent current in those systems. It is also important to realize that every geometrical dimension of the gold rings in the experiments was very much larger than the electronic screening length, and that the plasma frequency of gold greatly exceeds the characteristic energy scales that appear in the physics of persistent currents. Thus from the point of view of standard many-body theory, one would expect the gold rings to behave as a classic normal metal, and there seems to be no clear physical basis for a giant many-body enhancement of the persistent current. In view of the above, although the search for a many-body explanation of the experiments of Chandrasekhar *et al* [11] is far from over,[26] it seems reasonable to begin considering other possibilities.

The purpose of this article is to demonstrate that some simple three-dimensional models exhibit the combination of large persistent currents and small conductances that has been observed experimentally in the gold rings. These models do not introduce any unconventional electron-electron interaction effects in order to obtain large persistent currents. They are based instead on well-known materials and transport properties of thin gold films.

The previous theoretical work has been on models in which *random impurity scattering* is the mechanism determining the electron elastic mean free path $l$, and limiting the size of the per-sistent current. However, the rings of Chandrasekhar et al[11] were fabricated from gold films grown on oxidized silicon, an amorphous substrate. Such films are polycrystalline, and experimental studies have shown[27,28] that grain boundary scattering dominates the electronic transport properties of thin polycrystalline gold films at low temperatures. Random impurity scattering and surface roughness scattering are less important. It will be shown below that certain simple models that include grain boundary scattering exhibit the combination of large persistent currents and small conductances that is observed experimentally, but is not found in the random defect models. The large difference between the results obtained from the grain boundary and random impurity models is due to the fact that the persistent current and conductance of a ring are *physically different* quantities and depend in different ways on the *topology* of the defect structure in the ring.

It should be stressed that the puzzle presented by the experiments of Chandrasekhar et al[11] involves not only the persistent current but also the conductance of the ring, and that both of these quantities must be calculated on the same footing if the experimental results are to be explained convincingly. However, most of the previous theoretical studies have been devoted to the question whether a many-body enhancement of the persistent current is possible; much less attention has been given to calculating the conductance. In the present work both the persistent current and the conductance are calculated numerically, and both calculations represent exact solutions of the models considered. The reason why the grain boundary scattering models with non-interacting electrons that are discussed in this work exhibit large persistent currents together with small conductances, while random defect models do not, is due as much to the different behavior of the conductance as to the different behavior of the persistent current in the two classes of models.

Grain boundary scattering is the dominant scattering mechanism at low temperatures in thin films of gold,[27,28] and some other metals.[28] It therefore needs to be considered in any quantitative explanation of the experiments of Chandrasekhar et al.[11] Although, as is shown in this article, grain boundary scattering offers a possible explanation of the experiments of Chandrasekhar et al,[11] whether this will in the end turn out to be the whole story is an open question. The reason is that the gold rings in question have grain boundaries that are widely separated on the atomic scale and this "diluteness" of the grain boundaries is very important. Three-dimensional metal rings with large numbers of widely spaced grain boundaries are difficult to simulate numerically because of the practical limitations of computers. *Ab initio* calculations of the electronic structure of such systems are at present out of the question. Thus the present study has been limited to very simple models with separable Hamiltonians describing rings with arbitrary numbers of grain boundaries, and to more realistic (but still simple) non-separable model Hamiltonians describing rings with small numbers of dilute grain boundaries. Some important unresolved issues are how many grain boundaries the rings of Chandrasekhar et al[11] actually contained, and if this number was large, whether the present results can be applied to that case. It seems probable that these and many other questions will need to be resolved experimentally, and several good experimental tests of the present explanation of the observations of Chandrasekhar et al[11] are suggested at the end of this paper.

It should be noted that the present paper addresses the problem of the *typical* persistent current in a *single* metal ring, the quantity measured experimentally by Chandrasekhar et al.[11] The problem of the *average* ring persistent current over a large *array* of rings that was measured by Lévy et al,[10] although closely related, is significantly different. This is because the persistent currents carried by different rings in the array can differ in sign, which results in strong cancellations in the average. The theories of the average ring persistent current in large arrays of rings that have been developed to date[29, 14] have assumed that the electrons are scattered only by random impurities. In view of the results for single rings that are presented in this article, theoretical studies of arrays of rings based on models with grain boundaries would clearly be of interest, but are beyond the scope of the present paper.

## 2. Grain Boundary Models and Method of Solution

Consider a thin, multi-channel ring threaded by a magnetic flux, described by a Hamiltonian[30]

$$H = -\sum_{j,k,l} \zeta t^{\alpha}_{jkl} a^{\dagger}_{jkl} a_{j+1\,kl} + t^{\beta}_{jkl} a^{\dagger}_{jkl} a_{j\,k+1\,l} + t^{\gamma}_{jkl} a^{\dagger}_{jkl} a_{j\,k\,l+1} + h.c.$$

where $\zeta = \exp(2\pi i \Phi/Z)$, $\Phi$ is the magnetic flux threading the ring in units of the flux quantum $h/e$. $Z$ is the circumference of the ring in units of the lattice parameter. $a^{\dagger}_{jkl}$ is the electron creation operator at site $jkl$; the spin index is suppressed. The first term in $H$ represents electron hopping in the azimuthal direction around the ring; the other two terms describe hopping in the two orthogonal directions (see left inset, Fig.1, for the hopping directions $\alpha$, $\beta$ and $\gamma$). For a defect-free ring, the hopping coefficients $t_{jkl}$ in $H$ are all taken to be equal, $t^i_{jkl} = t_0$. Grain boundaries are modelled by setting those $t^i_{jkl}$ that represent hopping across a grain boundary equal to $t_{gb}$, with $t_{gb} < t_0$.

This is clearly a greatly simplified model of a metal ring with grain boundaries. However, it allows one to examine for the first time the effect on the persistent current of *extended* defects that cross the ring and *partition* it into "grains," as opposed to random impurities which do not. This *topological* difference between random impurity and grain boundary models is crucial, as will be shown below. Another important advantage of this model is that it can be solved numerically for both the persistent current and the conductance without making approximations. Thus all of the numerical results presented in this paper constitute *exact* solutions of the model.

The persistent current $I$ was evaluated at zero temperature by finding the eigenvalues of $H$ numerically, and using the result[2]

$$I = -e/h \; \partial E/\partial\Phi,$$

where $E$ is the total electronic ground state energy of the ring.

In addition to the persistent current, it is essential to calculate the resistance $R$ of the ring, since the value of the key experimental parameter $l/L$ was determined from resistance measurements. For definiteness, let $R$ be the resistance (in the absence of magnetic fields) of the wire made by severing the ring; the length of the wire is equal to the circumference $L$ of the ring. In this work the resistance $R$ is evaluated from the Landauer conductance formula

$$G = 2e^2 T/h,$$

where $R=1/G$ and $T =\mathrm{Tr}(\boldsymbol{t}\boldsymbol{t}^{\dagger})$ is the total multi-channel transmission probability through the wire at the Fermi energy.[31−34] On the other hand, in terms of the effective mean free path, the conductance is

$$G = 8\eta e^2 l/(3hL),$$

where $\eta$ is the number of conducting channels at the Fermi energy. Comparing these expressions for the conductance yields $l/L = 3T/4\eta$. Thus, in the present work,

$$\lambda \equiv 3T/4\eta$$

will represent the quantity $l/L$ which was inferred experimentally [11] from measured resistances. The parameter $T/\eta$ that appears in $\lambda$ has a simple physical meaning: it is the probability that an electron at the Fermi energy is transmitted right through the wire made by severing the ring. In this formulation, the puzzle raised by the experiments of Chandrasekhar *et al* [11] and to be addressed here is why gold rings with low electron transmission probabilities (values of $\lambda \sim 0.01$) are observed to display large persistent currents, of order $I_0$.

In this work, $T/\eta$, and hence $\lambda$, were calculated numerically by solving the Lippmann-Schwinger equation for the wire using a straight-forward generalization to three dimensions of the method of Nonoyama *et al*.[35]

The persistent current $I$ is periodic in the flux threading the ring, fluctuates as the number of electrons in the ring is varied, and depends on the precise configuration of the defects.[2-9] All of this must be considered when specifying what is meant by the "typical" persistent current $I_{typ}$. The persistent current may be written as a Fourier series in the magnetic flux:

$$I = \Sigma_{n=1}^{\infty} I_n \sin(2\pi n\Phi)$$

In the present work, $I_1$, the first Fourier coefficient, was calculated at zero Kelvin as a function of the number of electrons present in the ring. Its root mean square value $I_{rms} \equiv \sqrt{<I_1^2>}$ was evaluated, the average $<...>$ taken over all electron populations ranging from 1/2 to 1 electron per site.[36] $I_{rms}$ was then averaged over configurations of grain boundaries. This configuration average will be referred to as $I_{typ}$, since it is a typical value of the persistent current in an isolated ring. Since $\lambda$ also fluctuates with the Fermi energy and defect configuration, $\lambda_{av}$, the average of $\lambda$ over grain boundary configurations and Fermi energy was calculated. For consistency with the above definition of $I_{typ}$, the range of Fermi energies over which $\lambda$ was averaged was $[-1.73t_0, 0]$; this corresponds to electron populations ranging from 1/2 to 1 electron per site of Hamiltonian $H$, for a bulk system.

## 3. Results for Separable Hamiltonians: Simple Grain Boundaries

Let us consider, to start with, rings with only "radial" grain boundaries (see left inset, Fig.1). In the Hamiltonian $H$, this means that for certain values of $j$ (those at grain boundaries) $t_{jkl}^{\alpha} = t_{gb}$ for all $k$ and $l$, while all other $t_{jkl}^{i}$ are equal to $t_0$.[37] Thus $H$ is separable. [Similar results were obtained for non-separable Hamiltonians (for rings with small numbers of dilute grain boundaries running in arbitrary directions with steps on the atomic scale, and for rings with small numbers of atomically-rough grain boundaries) and will be discussed in Section 4.] The rings simulated had square wire cross-sections of $N \times N$ sites (see left inset, Fig.1), and aspect ratios $M \equiv Z/N$, for a ring circumference of $Z$ sites.

In typical thin gold films the grain size is of roughly equal to the thickness of the film[27,28] which would imply that the number of grain boundaries in the ring is roughly equal to $M$, the aspect ratio. However the grain size can be larger for some substrates and/or after heat treatment.[27] The number of grain boundaries in the gold rings of Chandrasekhar *et al*[11] was not known accurately, but may have been significantly smaller than $M$.[38] But, in the absence of more detailed information, it seems reasonable to examine closely models where the number of grain boundaries equals $M$, choosing the grain boundary strength parameter $t_{gb}/t_0$ in such a way as to make the transmission probability $\lambda_{av}$ of the ring approximate the experimental value, $\lambda_{av} \sim 0.01$. This choice means that in a ring with more grain boundaries, the reflection probability at each individual grain boundary is set to be weaker than in a ring with fewer grain boundaries so that the total transmission probability of an electron around the ring is approximately the same.

This is the case considered in Fig.1 where the symbols $\circ$, $\square$, and $\blacktriangle$ show $I_{typ}/I_0$ [39] and $\lambda_{av}$ as a function of the wire cross-section $N$, computed for three choices of the aspect ratio $M$, for radial grain boundaries. In each case the number of grain boundaries in the ring is equal to $M$. The configuration averages were computed assuming grain boundaries randomly distributed around the ring. In each case, the value of $t_{gb}$ was chosen so that $\lambda_{av} \sim 0.01$ ($t_{gb}/t_0 = 0.066$, 0.210 and 0.485 for $M = 1$, 4 and 16, respectively); see the lower right inset of Fig.1 for the precise values of $\lambda_{av}$ computed for these values of the grain boundary parameters. The straight lines in Fig.1 are guides to the eye.

Notice that for N~30, $I_{typ} \sim I_0$ even though $\lambda_{av} \sim 0.01$. This demonstrates that a model with non-interacting electrons can exhibit a large persistent current $\sim I_0$ despite the presence of scattering that results in a short transport mean free path $l \sim 0.01L$ and correspondingly small conductance for the ring. Note also that in Fig.1, for fixed $M$, $I_{typ}/I_0$ increases *linearly* with $N$ (there are small fluctuations about the straight line behavior), while $\lambda_{av}$ slowly decreases. This behavior is in marked contrast with the case of random impurity scattering where $I_{typ}$ is predicted[9] to be two orders of magnitude smaller ($I_{typ}/I_0 \sim \lambda$) for the same value of the conductance, and $I_{typ}/I_0$ is predicted to be independent of $N$.

In Fig.1, $I_{typ}/I_0$ also increases when the aspect ratio $M$ (and number of grain boundaries) increases at fixed $\lambda_{av}$, but the change of $I_{typ}$ from $M = 4$ to $M = 16$ is small. The fact that $I_{typ}/I_0$ increases somewhat while the number of grain boundaries is increasing is not unreasonable since the scattering strength of the individual grain boundaries is decreasing so as to keep the transmission probability $\lambda_{av}$ of the whole ring approximately fixed.

Since $\eta$, the number of transverse modes of the ring, is proportional to $N^2$, the linear dependence of $I_{typ}/I_0$ on $N$ in Fig.1 resembles the result $I_{typ} \sim I_0\sqrt{\eta}$ predicted[9] for ballistic (defect-free) rings. However here $I_{typ} << I_0\sqrt{\eta}$. Also, unlike the ballistic case, here one must vary not just $N$ (which controls $\eta$) but also the circumference $Z$ (keeping $M = Z/N$ constant) to see linear behavior. This is shown in Fig.2, where $I_{typ}/I_0$ is plotted for rings with 4 radial grain boundaries ($t_{gb}/t_0 = 0.210$), for $M = 4$ and 16, and also for fixed $Z = 12$. Increasing $N$ (and $\eta$) at fixed $Z$ results in slower growth of $I_{typ}/I_0$ than the linear behavior seen for fixed $M = 4$ and 16. Thus the channel counting arguments that yield the $\sqrt{\eta}$ scaling of $I_{typ}/I_0$ in *ballistic* rings are not adequate for rings with grain boundaries. For the latter, the aspect ratio $M$ is also significant; the behavior of $I_{typ}/I_0$ at very small values of $M$ (where the *concentration* of grain boundaries in the ring is relatively high) is different from what it is at high M where the grain boundaries are more dilute. Notice also that $I_{typ}/I_0$ changes little when $M$ changes from 4 to 16 in Fig.2, or when the number of grain boundaries changes from 4 to 16 at fixed $\lambda_{av}$ for $M = 16$ (compare Fig.1 and Fig.2).

It is of interest also to consider separately the dependence of the persistent current on the aspect ratio of the ring and on the number of grain boundaries. This is addressed in Fig.3 and Fig.4. Fig.3 shows examples of the dependence of $I_{typ}/I_0$ on the aspect ratio $M$ for a fixed wire cross-section $N = 20$ and for fixed numbers of grain boundaries. The results for 4 grain boundaries with $t_{gb}/t_0 = 0.210$ and for 16 grain boundaries with $t_{gb}/t_0 = 0.485$ are shown. (These grain boundary "strengths" are the same as for the $M=4$ and $M=16$ results shown in Fig.1, respectively). For small values of $M$ the dependence of $I_{typ}/I_0$ on the aspect ratio $M$ is non-monotonic and qualitatively different in the two cases shown. But at larger values of $M$, $I_{typ}/I_0$ becomes almost independent of $M$, in agreement with the results shown in Fig.2. The fact that the behavior seen for very low values of $M$ in Fig.3 is anomalous should not be surprising since there the grain boundaries themselves constitute a large fraction of the material of the ring and *grossly* alter its electronic structure.

In Fig.4, ●, ▲ and ■ show $I_{typ}/I_0$ as function of the number of grain boundaries for a fixed aspect ratio $M=4$ and wire cross-section $N=20$, for $t_{gb}/t_0 = 0.800$, 0.485 and 0.210 respectively. Notice that here in each case the strength the individual grain boundaries is not changed as the number of grain boundaries in the ring increases (in contrast to Fig.1). In Fig.4, for small numbers of grain boundaries, the persistent current begins to decrease exponentially with increasing number of grain boundaries. As the number of grain boundaries continues to increase and the grain boundaries cease to be dilute, $I_{typ}/I_0$ begins to deviate from the exponential behavior. This deviation is strongest for the more strongly scattering grain boundaries. For weak scattering the exponential behavior persists to significantly higher numbers of grain boundaries in the ring. That the breakdown of the exponential behavior is principally an effect of the increasing concentration of grain boundaries in the ring can be seen by considering the results denoted by ◇ in Fig.4. These again show $I_{typ}/I_0$ for $N=20$ and $t_{gb}/t_0 = 0.210$ (as for the results denoted ■), but in this case $M$ is not fixed but set equal to the number of grain boundaries, so that the *concentration and the strength* the of grain boundaries in the ring are *both* kept fixed when the number of grain boundaries increases. I.e. here the number of grain boundaries is proportional to the circumference of the ring. In this case it can be seen that dependence of $I_{typ}/I_0$ on the number of grain boundaries is closer to exponential, when the number of boundaries (and the ring circumference) are large.

If the number of grain boundaries in the ring exceeds the aspect ratio $M$, the linear dependence of $I_{typ}/I_0$ on $N$ at constant $M$ that is seen in Fig.1 breaks down for small $N$. This is illustrated in the inset of Fig.2 where $I_{typ}/I_0$ is plotted against $N$ for 4 radial grain boundaries with $t_{gb}/t_0 = 0.210$ and a fixed aspect ratio $M = 1$. The breakdown of the linear behavior at low $N$ occurs because if the aspect ratio and the number of grain boundaries are both kept fixed, the grain boundaries themselves take up an increasingly large fraction of the material of the ring as $N$ decreases. As has already been seen for low aspect ratios $M$ in Fig.3, this changes the underlying electronic structure of the ring very strongly and modifies the behavior of the persistent current. Clearly the gold rings of Chandrasekhar *et al*,[11] where the grain boundaries most likely comprise less than 0.1% of the material of the ring, are in a dilute regime where this effect does not occur. But it becomes a major concern when modelling rings with non-separable Hamiltonians, since only quite small systems of that type can be simulated numerically, which severely limits the number of grain boundaries in rings that can be modelled if the grain boundary concentration is to be kept low.

## 4. Results for Non-Separable Hamiltonians

The above results are for Hamiltonians with radial grain boundaries. It is necessary to establish whether they also apply to systems with grain boundaries running in *arbitrary* directions. The Hamiltonians of such systems are non-separable,[40] and their numerical study is much more difficult, requiring the use of Lanczos algorithms[40] to calculate the electronic energy spectra. Some representative results are given in Fig.5. Here, for the results denoted ● and ○, the number of grain boundaries is equal to the aspect ratio $M$, as in Fig.1, and $t_{gb}/t_0$ is chosen so that $\lambda_{av} \sim 0.01$. The grain boundaries considered are planes slicing all the way through the ring, as is illustrated in the right inset of Fig.5. Planar grain boundaries are realistic, being favored by free-energy considerations during sample preparation. But in the simulations the "planes" were in fact stepped interfaces on the atomic scale of the lattice Hamiltonian $H$. Configuration averaging was performed over the possible orientations of such grain boundaries, but *excluding* boundaries parallel to the β or γ (or both the β and γ) hopping directions in the Hamiltonian. This exclusion ensured that averaging was *only* over non-separable Hamiltonians. These results in Fig.5 are similar to those in Fig.1: $\lambda_{av}$ slowly decreases with increasing $N$. $I_{typ}/I_0$ increases linearly with $N$ for one grain boundary with $M$ =1 (●). For two grain boundaries with $M$ =2 (○), the behavior is also consistent with linear growth of $I_{typ}/I_0$ with $N$, but the fluctuations are larger, and the accessible values of $N$ smaller. (The fluctuations are a finite size effect; they do not disappear when more grain boundary configurations are included in the average $I_{typ}$.) These results indicate that the behavior described above for radial grain boundaries is not specific to separable Hamiltonians. The most obvious difference between the above results for the systems with radial and non-radial grain boundaries is that in the latter case $I_{typ}$ is smaller by a factor ~ 2, for the same value of $\lambda_{av}$.

A feature of the results for 2 grain boundaries in the $M$=2 structure that are denoted ○ in Fig.5, is that at low $N$ where the concentration of grain boundaries in the crystal is highest, $I_{typ}/I_0$ is almost independent of $N$ (for $N \leq 6$). A similar but less pronounced tendency for $I_{typ}/I_0$ to level out at low $N$ can also be discerned in the results ● for the $M$=1 structure with one grain boundary. If the grain boundaries in the ring are made more concentrated, this tendency of $I_{typ}/I_0$ to level out at low N becomes stronger. This is illustrated by the results denoted ◆ in Fig.5 which show $I_{typ}/I_0$ for 2 grain boundaries in an aspect ratio $M$ =1 structure, with $t_{gb}/t_0 = 0.076$. Here $I_{typ}/I_0$ is almost independent of $N$ for $N$<15, and then shows an increase, but it is not clear whether the linear growth of $I_{typ}/I_0$ with $N$ is recovered at large $N$ or not, because of the fluctuations that $I_{typ}/I_0$ shows as a function of $N$ and because of the limited range of $N$ that is accessible numerically.

Notice that the breakdown of the linearity of $I_{typ}/I_0$ with $N$ over the *wide* range of $N$ seen for the 2 grain boundary $M$=1 system (◆) does not occur in the 2 grain boundary M=2 system (○) or in the 1 grain boundary $M$=1 system (●). Thus neither the number of grain boundaries alone nor the value of $M$ alone controls this effect. An important factor appears to be the grain boundary *concentration*. These results (like those presented in Fig.2, Fig.3 and Fig.4, and the discussion at the end of Section 3) demonstrate the importance of simulating systems in which the grain boundaries are *dilute*, as they are in the real gold rings. However, for models with larger values of the aspect ratio $M$, the numerically accessible range of $N$ shrinks very rapidly. This, together with the fluctuations that $I_{typ}/I_0$ shows as a function of $N$, makes it very difficult to establish how $I_{typ}/I_0$ scales with $N$ for large numbers of dilute non-radial grain boundaries.

Although the grain boundaries discussed above correspond to non-separable Hamiltonians, are stepped on the atomic scale and are randomly located and oriented in the ring, each of them has a perfectly regular microscopic structure (except where two grain boundaries cross). Recent high resolution electron microscopy studies[41] have yielded very detailed and clear images of grain boundaries in gold that are perfectly ordered on the atomic scale. However, earlier work [42] indicated some grain boundaries in gold to be very well ordered, but others to exhibit some atomic scale disorder in a region along the grain boundary a few atoms thick. Since the microstructure of the grain boundaries of the gold rings of Chandrasekhar *et al.*[11] was not measured, it is unclear which type of grain boundary was more typical, and it is of interest to simulate atomically rough grain boundaries as well.

The results of simulations for a model of a ring with an atomically-rough grain boundary, are shown by the + symbols in Fig.1 for $M$ = 2. This grain boundary contains a central layer identical to one of the radial grain boundaries described in Section 3, but surrounded by two disordered layers. The disordered layers are constructed by making 50% of the hopping matrix elements in the α direction (at randomly chosen sites) in the lattice layers adjacent to that of the central layer also weak and equal to $t_{gb}$. The results for this Hamiltonian are again similar to those for the simple radial grain boundaries shown in Fig.1. The effect on the (configuration averaged r.m.s.) persistent current of roughening the grain boundary in this way is similar to that of tilting the grain boundary, i.e., making it non-radial.

## 5. Comparison of Grain Boundary and Random Defect Models

Why do these grain-boundary models yield the combination of large persistent currents and small conductances that is observed experimentally, while random defect models do not? This is addressed in Fig.6. Here $I_{typ}/I_0$ and $\lambda_{av}$ are results for $M = 1$ (one radial grain boundary) taken from Fig.1. $I_r$ and $\lambda_r$ are the persistent current and transmission defined in the same way as $I_{typ}$ and $\lambda_{av}$, but with the same "defects" (terms in $H$ for which $t_{jkl}^{\alpha} = t_{gb}$) scattered randomly through the system instead of forming a continuous grain boundary. The striking feature is that while $\lambda_{av}$ is smaller than $\lambda_r$ by a factor ~40, $I_{typ}$ is smaller than $I_r$ by only a factor ~2. That is, collecting the random defects into a coherent grain boundary forming a barrier extending all the way across the path of the current, reduced the transmission $\lambda$ (conductance of the ring) far more than it did the persistent current.

This clarifies why large persistent currents occur for low conductances in the grain boundary model, but not in random defect models. The physics underlying this numerical example is that the mean free path is a *transport* property, while the persistent current is an *equilibrium* effect. The persistent current is given by the flux-derivative of the total electronic energy of the ring.[2] The total energy (and hence the persistent current) is not very sensitive to the arrangement of the defects, and not disturbed much if defects form into a grain boundary, i.e., a continuous "barrier" across the ring. But transport (the conductance or $\lambda$) is well known to be *greatly* affected by such a barrier. Analogous (and more extreme) differences between the behavior of transport coefficients and equilibrium quantities are commonly associated with percolation phenomena where the topology of the sample is crucial.

This physical argument suggests that the present results should not be very sensitive to the shape of the grain boundary, so long as it forms a continuous barrier across the ring. For example, as has been shown above for a ring with one grain boundary, making the grain boundary rough on the atomic scale does not change the predictions of the model qualitatively.

It should be stressed that it is the conductance rather than the persistent current that is sensitive to the arrangement of defects in Fig.6. Thus the present work suggests that the explanation of the experiments of Chandrasekhar *et al.*[11] has at least as much to do with the behavior of the conductance of the ring as with that of the persistent current, and that the conductance aspect of the problem deserves much more attention than it has received in previous theoretical studies.

It is also clear from the above discussion of Fig.6 that, in order to achieve a low electron transmission probability ($\lambda \sim 0.01$), the random impurity model requires a much larger number of *atomic* defects to be present in the ring than is required in the grain boundary model (if the individual atomic defects are similar). This in turn depresses the value of the persistent current that the random impurity model predicts for a given conductance. Since the resistance of gold films is mainly due to scattering by grain boundaries,[27,28] the high concentration of defects that is required by random impurity models in order to obtain low conductances is a deficiency of those models. That is, the random defect models assume the gold rings to be much "dirtier" than they really are. For example, computer simulations of persistent currents in rings with random impurities have tra-ditionally employed Anderson models in which *every* site is an impurity site and the spread of random site energies $W$ is not very much smaller than the tight binding energy band parameter $t_0$.

## 6. Comparison with Experiment

The amplitudes of the persistent currents measured by Chandrasekhar *et al* [11] ranged in magnitude from $0.3I_0$ to $2.0I_0$ over the three rings for which results were reported. For these rings, $N \sim 300$, so that the largest wire cross-sections $N$ considered in Fig.1 and Fig.5 are about an order of magnitude smaller than in the experiments. The aspect ratios of the experimental rings were in the range $M \sim 100$ to $200$. The numbers of grain boundaries in the rings were unknown, but the transmission probabilities of the rings could be estimated from the known conductances of similar samples to yield $\lambda \sim 0.01$, as discussed in Section 2. If one assumes that the rings contain any given number of grain boundaries, one can choose the grain boundary model parameter $t_{gb}/t_0$ in such a way that the electron transmission probability of the model ring matches that in the experiments, and this was done in the calculations presented in Fig.1 and Fig.5.

One can extrapolate the results of the simulations reported in Section 3 for rings with radial grain boundaries to systems of with the experimental values of $N$ and $M$, by using the linearity of $I_{typ}/I_0$ with $N$ demonstrated in Fig.1 and the fact that for fixed $N$ and fixed numbers of grain boundaries, $I_{typ}/I_0$ is independent of $M$ at large $M$, as shown in Fig.3. This yields $I_{typ} \sim 10\,I_0$ for rings with the experimental dimensions, if the rings contain 4 or 16 grain boundaries and the ring conductance matches experiment, i.e. $\lambda \sim 0.01$. It should be noted that in the models with radial grain boundaries, for large numbers of grain boundaries and large $M$, $I_{typ}/I_0$ is insensitive to both $M$ and the number of grain boundaries *if $\lambda$ is held fixed*. Thus the prediction $I_{typ} \sim 10\,I_0$ for the radial grain boundary models is insensitive to the number of grain boundaries that one assumes the ring to contain, if the ring's N, $M$ and conductance are required to match their experimental values.

For non-radial grain boundaries, one can extrapolate the linear behavior of the results seen in Fig.5 for $M = 1$ structures with one grain boundary, and for M=2 structures with two grain boundaries, to the experimental value of $N$, *while holding $M$ fixed*. This yields $I_{typ} \sim 3I_0$ if $N$ and the conductance of the ring take the experimental values. For 2 grain boundaries, the results of Fig.5 show that $I_{typ}/I_0$ *increases* when $M$ increases from 1 to 2. Thus it is not unreasonable to expect rings with one or two non-radial grain boundaries, and having the experimental values of $N$, $M$ and the conductance, to exhibit large values of the persistent current $I_{typ} \gtrsim 3I_0$, that are quite similar to those observed by Chandrasekhar *et al.*[11] However, as discussed in Section 4, for large numbers of non-radial grain boundaries the important dilute regime (of large $N$ *and* $M$) is not readily accessible to numerical investigation; a definitive treatment of that regime is beyond the scope of the present work.

It should also be emphasized that both the separable and non-separable Hamiltonians considered in this article represent very simple models of grain boundaries. Complete *ab initio* calculations of more realistic grain boundary structures, of the associated electronic eigenstates and of the persistent currents and transport in rings containing them would clearly be of interest.

To summarize: There is good experimental evidence that grain boundary scattering is the dominant electronic scattering mechanism in thin gold films.[27,28] The present simulations have demonstrated that exact solutions of some simple models that include grain boundary scattering exhibit the combination of large persistent currents and small conductances that is observed experimentally in gold rings,[11] but cannot be explained by random impurity models with non-interacting electrons. The random impurity models yield persistent currents that are too small by factors of 30 to 150.[11] However, it is unclear how many grain boundaries the gold rings of Chandrasekhar *et al.*[11] actually contained, and whether more realistic models or models with many non-radial grain boundaries can account for the experimental data as well as the simple models discussed above do, is yet to be determined.

## 7. Proposed Experimental Tests

It is clear that direct experimental tests of the predictions of the grain boundary scattering explanation of the persistent current measurements of Chandrasekhar *et al.*[11] are needed, and some are suggested below:

1) The experimental data reported by Chandrasekhar *et al.*[11] exhibited a sample-to-sample variation of the size of the observed persistent current ($I/I_0$) by a factor of ~ 7, over the three samples. This is consistent with the idea that scattering by grain boundaries (whose number varies randomly from sample to sample) controls the magnitude of the persistent current; see, for example, Fig.4. However, similar variations in the magnitude of the persistent current are also found to occur in simulations if the number of grain boundaries is kept fixed but their *configuration* changes. (Recall that the numerical values of $I_{typ}$ presented in this paper are grain boundary configuration averages of r.m.s. values, as defined in Section 2.) It would be of interest to establish experimentally whether there is a correlation between the magnitude of the persistent current and the number of grain boundaries that the sample contains, by performing persistent current measurements followed by spatially resolved electron diffraction measurements on a series of samples. But in such a study it would be important to bear in mind the sensitivity of the persistent current to the details of the grain boundary configuration; comparisons between a small number of samples with similar numbers of grain boundaries would not be significant. It would also be very desirable to measure the conductances of the same samples directly instead of relying on conductance measurements on similar samples, as has been done in the past.

2) It should be possible to fabricate *single crystal* gold rings that do not contain any grain boundaries at all. These should exhibit significantly larger persistent currents than those that have been observed to date.

3) If single crystal rings of gold *alloys* can be fabricated, these should exhibit *weak* persistent currents for low conductances, since the random impurity theories[9,16] should apply to such systems.

4) Semiconductor rings such as those of Mailly *et al.*[12] do not contain any crystal grain boundaries in the conducting region. The quasi-ballistic rings of this type that have been studied to date have shown no evidence of any anomalous behavior; the magnitudes of the persistent currents observed in them are consistent with the predictions of standard, non-interacting electron theories, which for this case predict that $I_{typ} \sim I_0$ since the number of channels is small and $l \gtrsim L$. By introducing impurities directly into the region occupied by the conducting electrons it should be possible to fabricate diffusive semiconductor rings for which $l << L$ but which are still free of grain boundaries. It would be very interesting to establish whether or not large persistent currents $\sim I_0$ occur in such samples. However, it is necessary to measure both the persistent currents and conductances of such samples for this test to be meaningful. It should also be stressed that a truly diffusive system, preferably with point defect type scatterers, is required for this test to be convincing -- Using *charged* random impurities with long-range interactions can result in a few electronic modes following a preferred path through the ring and thus yield misleading results.

5) Although the semiconductor rings are free of grain boundaries, it is possible to impose an electrostatic potential barrier (the analog of a grain boundary) across such a ring by means of a gate. In such a system it should be possible to measure both the transmission probability of electrons through the barrier and the persistent current in the ring. It is predicted[43] that in the regime of quantum tunneling of the Fermi electrons just below the top of the barrier, large persistent currents $\sim I_0$ can coexist with small electron transmission probabilities $<< 1$. It would be of interest to observe this semiconductor analog of the large persistent currents that coexist with low conductances in the small gold rings.

I wish to thank C. Barnes, B. Heinrich, B. L. Johnson, D. Loss and R. A. Webb for interesting discussions. This work was supported by the Natural Sciences and Engineering Research Council of Canada.

# Figure Captions

**Fig.1** Normalized persistent current $I_{typ}/I_0$ and average transmission $\lambda_{av}$ vs. $N$, for radial grain boundaries, are shown as ○, □, and ▲ for ring aspect ratios $M =$1, 4 and 16 respectively. Number of grain boundaries $= M$. Straight lines are guides to the eye. + symbols show results for an *atomically rough* grain boundary, for $M$=2 and $t_{gb}/t_0 = 0.047$. Left inset: Schematic of ring with 4 radial grain boundaries (shaded).

**Fig.2** $I_{typ}/I_0$ vs. $N$ for 4 radial grain boundaries. □, +, ◆ are results for fixed aspect ratios $M = 4$ and 16, and a fixed circumference $Z =$12, respectively. Inset: $I_{typ}/I_0$ vs. $N$ for 4 radial grain boundaries with $t_{gb}/t_0 = 0.210$ and a fixed aspect ratio $M =$1. Note the breakdown of linearity of $I_{typ}/I_0$ with $N$ at low $N$ where the grain boundaries themselves constitute a significant fraction of the material of the ring.

**Fig.3** $I_{typ}/I_0$ vs. aspect ratio $M$ for a fixed wire cross-section $N = 20$ and for fixed numbers 4 and 16 of radial grain boundaries.

**Fig.4** ●, ▲ and ■ show $I_{typ}/I_0$ vs. the number of radial grain boundaries for a fixed aspect ratio $M$=4 and wire cross-section $N$=20, for $t_{gb}/t_0 = 0.800$, 0.485 and 0.210 respectively. ◇ also show $I_{typ}/I_0$ for $N$=20 and $t_{gb}/t_0 = 0.210$ (as for the results denoted ■), but with the aspect ratio $M$ set equal to the number of grain boundaries, so that the *concentration* of grain boundaries in the ring is kept fixed. The straight lines are guides to the eye.

**Fig.5** $I_{typ}/I_0$ and $\lambda_{av}$ vs. $N$, for non-radial grain boundaries. ● and ○ are results with the number of grain boundaries equal to the aspect ration $M$, for aspect ratios $M =$1 and 2 and $t_{gb}/t_0 = 0.060$ and 0.075 respectively. The results denoted ◆ show $I_{typ}/I_0$ for 2 grain boundaries in an aspect ratio $M$ =1 structure, with $t_{gb}/t_0 = 0.076$. Error bars are the statistical uncertainty due to averaging over finite numbers of grain boundary configurations. Straight lines are guides to the eye. Right inset: Schematic of 2 non-radial grain boundaries.

**Fig.6** Comparison of random defect and grain boundary models. See text.